\documentclass[12pt]{article}
\usepackage{amssymb,amsmath,latexsym,amsthm,amsfonts}
\usepackage[english]{babel}
\usepackage[T1]{fontenc}
\usepackage{color}
\usepackage{graphics}
\usepackage{graphicx}

\usepackage[colorinlistoftodos]{todonotes}

\newtheorem{theorem}{Theorem}[section]

\newtheorem{proposition}[theorem]{Proposition}

\def\1g{1\hskip -3pt \mbox{l}}

\small\normalsize

\title{Investigating linear relationships between non constant variances of economic variables
}

\author{
{\sc Junichi Hirukawa$^{a}$, Hamdi Ra\"{\i}ssi$^{b}$\footnote{Instituto de Estadistica, Pontificia Universidad Cat\'{o}lica de Valpara\'{\i}so, Err\'{a}zuriz 2734, Valpara\'{\i}so, Chile. E-mail: hamdi.raissi@pucv.cl. This paper was supported by CONICYT-FONDECYT under grant no. 1160527.}
\footnote{We thank two anonymous reviewers and the associate editor whose comments helped to improve the paper.}
}}

\begin{document}

\maketitle
\begin{center}
$^a$ Department of Mathematics,
Faculty of Science,
Niigata University, Japan.\\$^b$ Instituto de Estadistica, Pontificia Universidad Cat\'{o}lica de Valparaiso, Chile.
\end{center}

\noindent {\em Abstract:} In this paper we aim to assess linear relationships between the non constant variances of economic variables. The proposed methodology is based on a bootstrap cumulative sum (CUSUM) test. Simulations suggest a good behavior of the test for sample sizes commonly encountered in practice. The tool we provide is intended to highlight relations or draw common patterns between economic variables through their non constant variances. The outputs of this paper is illustrated considering U.S. regional data.

\vspace*{.4cm} \noindent {\em Keywords:} Common variance patterns; VAR models; CUSUM tests; wild bootstrap.\\

\vspace*{.4cm} \noindent {\em JEL codes:} C12; C32; R10.

\section {Introduction}
\label{intro}

In time series econometrics it is a common approach to study links between variables with constant covariances by establishing long run relationships. A classical example for this task is given by the cointegration concept introduced in Engle and Granger (1987) (see Johansen and Nielsen (2019) for recent developments in the field). Reference can be also made to Engle, Ng and Rothschild (1990) who introduced factor GARCH models to study co-movements in the volatilities of returns variables. M\"{u}ller and Watson (2018) highlight long-run covariability for various U.S. economic variables. In the literature extensions of these concepts in the case of processes with non stationary covariance are available. For instance Kim and Park (2010), Cavaliere, Rahbek and Taylor (2010) and Wang, Zhao and Li (2019) investigated the statistical analysis of cointegrated systems with non constant unconditional covariance. Cardinali and Nason (2010) introduced the concept of costationarity. More precisely they supposed the possible existence of linear combinations of locally stationary processes producing a stationary process. This interest for time series with non constant variance may be explained by the fact that such feature is commonly encountered in practice. For instance Sensier and van Dijk (2004) considered 214 U.S. macroeconomic variables and found that approximately 80\% of these variables have a changing variance through time. 

In this paper we wish to test the existence of a linear relationship between the non constant variance of economic variables. The proposed tools are intended to complement the use of the above cited concepts in applied studies, by establishing common behaviors between variables through their non constant variance. Our methodology is based on the residuals obtained from the estimation of vector autoregressive (VAR) model for the conditional mean of the observed process. A natural way to detect departures from a linear relation between variance structures is to consider a cumulative sum (CUSUM) statistic. However using a result of Hansen (1992a), it is found that the test statistic has a non standard asymptotic behavior under the null hypothesis. As a consequence we use the wild bootstrap method to implement our test (see Wu (1986) for the wild bootstrap method).

\begin{figure}[h]
\begin{center}
\includegraphics[scale=0.57]{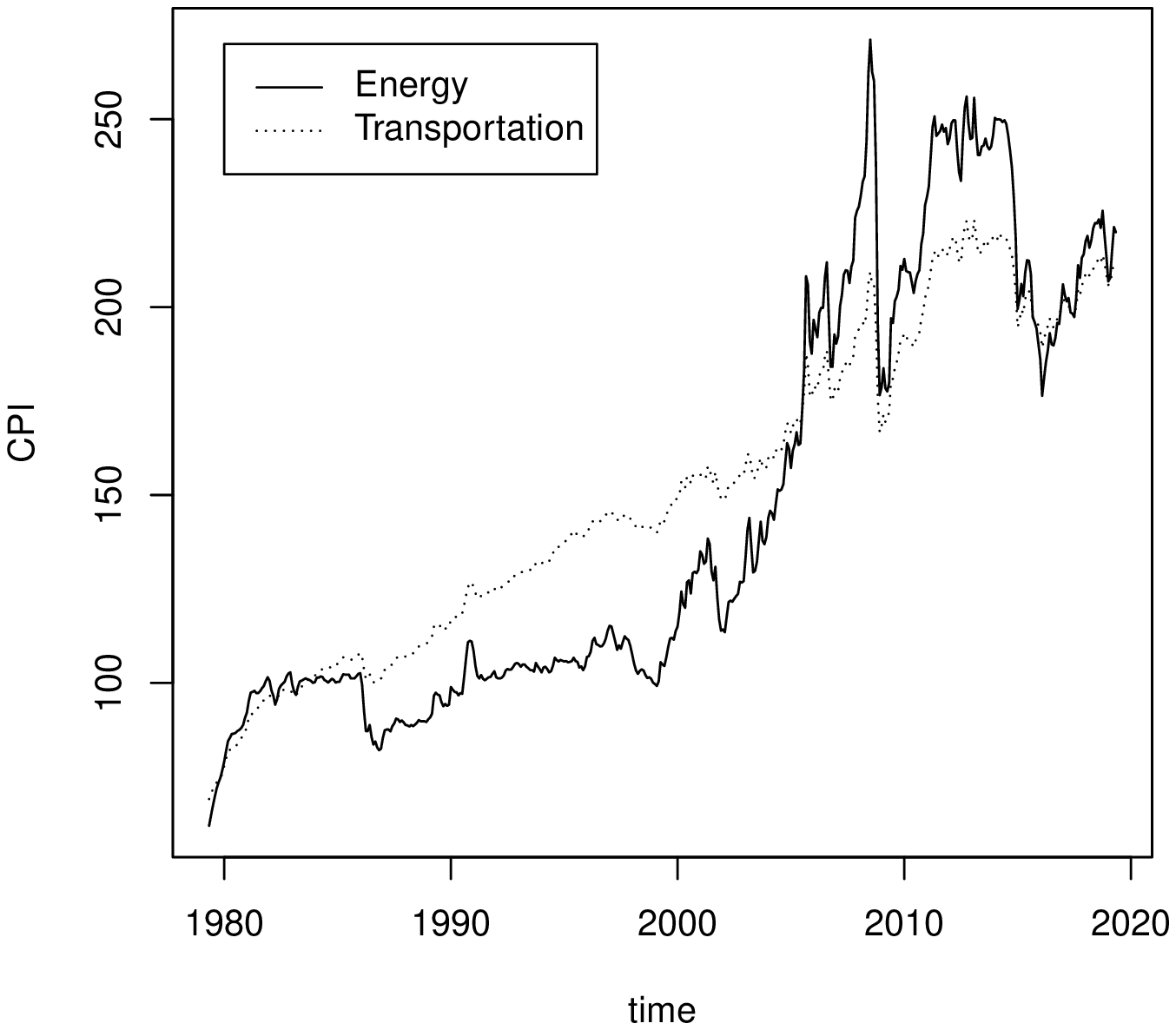}
  \end{center}
\caption{Crossplot of the squared first component residuals (on the $x$-axis) and the squared second component residuals (on the $y$-axis) of the transportation-energy CPI VAR modelling.}
\label{one}
\end{figure}

\begin{figure}[h]
\begin{center}
\includegraphics[scale=0.42]{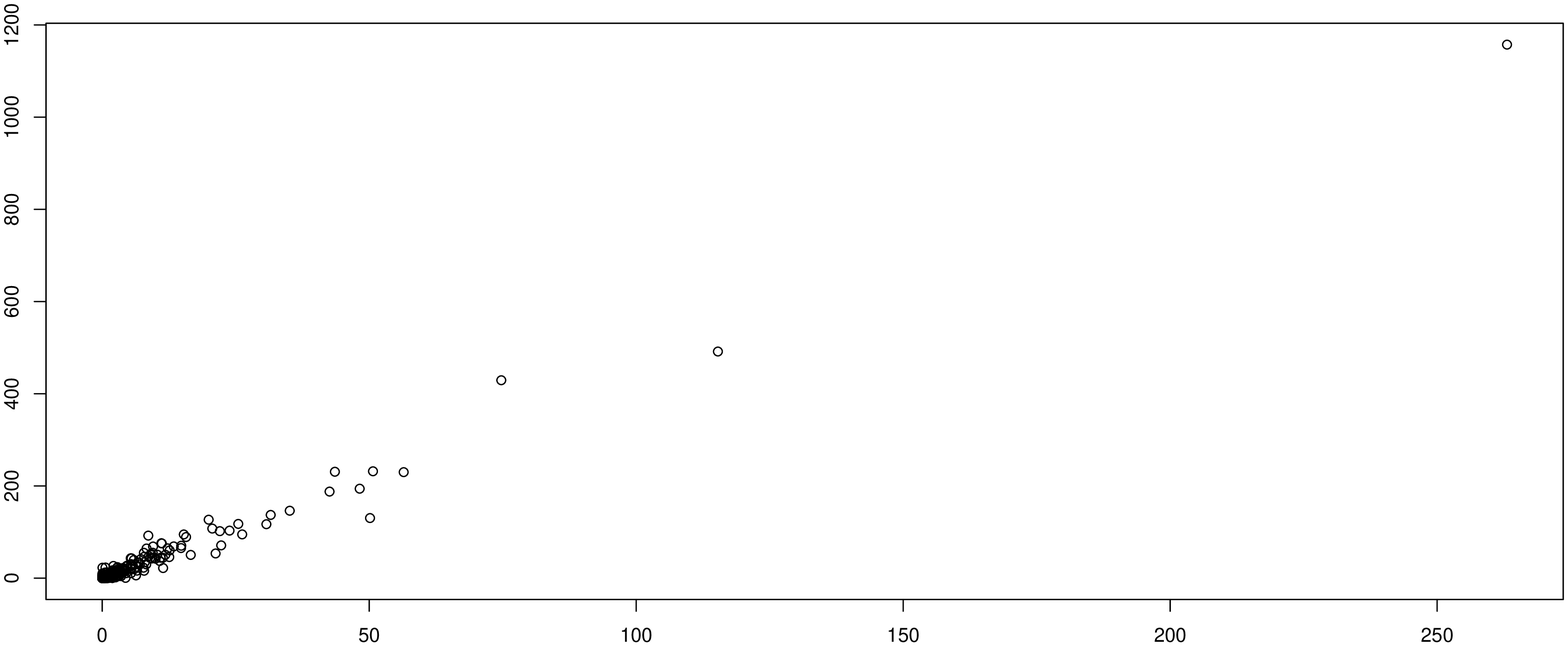}
  \end{center}
\caption{Crossplot of the squared first component residuals (on the $x$-axis) and the squared second component residuals (on the $y$-axis) of the transportation-energy CPI VAR modelling.}
\label{cross}
\end{figure}

The tool developed in this paper can be useful for a variety of econometric studies.
For instance it can be used to form clusters of seasons by determining common heteroscedastic patterns (see Trimbur and Bell (2010) for more details on seasonal heteroscedasticity). For regional data it can be considered to test if the amplitude changes are shared by sets of individuals (see the Consumer Price Indexes (CPI) of U.S. cities example in Section \ref{realdata} below). More generally it allows to investigate how the variability of a given variable can impact the variability of another one, as illustrated in the following example. Let us consider the monthly U.S. energy and transportation consumer price indexes (CPI) for all urban consumers from May 1, 1979 to May 1, 2019 (see Figure \ref{one}).\footnote{The data may be downloaded from the website of the research division of the Federal Bank of Saint Louis. https://fred.stlouisfed.org/} The levels and dynamics of such a kind of data are often studied using cointegration and causality concepts (see among others Oh and Lee (2004), Lise and Van Montfort (2007), Chang \textit{et al.} (2009) or Ramos and Veiga (2014)). However these contributions may be complemented by identifying common time-varying variance behaviors. Following our methodology we begin by filtering the conditional mean of the first differences of the variables using a VAR model. Now examining the squared residuals obtained from the previous step in Figure \ref{cross}, it is reasonable to suppose a proportional relationship between the CPI variances. From this observation, it turns out that the variability of the transportation CPI is driven by the variability of the energy CPI. More generally such kind of study may help econometricians to identify the sources of variable's time-varying instability.


The structure of this paper is as follows. In Section \ref{S1} we first set the dynamics ruling the observed process. In a second step we introduce the CUSUM test statistic and derive its asymptotic properties. The bootstrap procedure is also introduced and some practical issues are discussed. In Section \ref{numsection} numerical experiments are conducted to shed some light on the finite sample behavior of the studied test. The use of the proposed methodology is illustrated using U.S. regional data. The proofs of the theoretical results are gathered in Section \ref{secproofs}.\\


\section{Testing for linear relations between non constant variance structures}
\label{S1}

Let us consider the bivariate VAR process $X_{t,T}$:

\begin{eqnarray}\label{model}
X_{t,T}=\mu_0+\sum_{j=1}^{p}A_{0j}Z_{t-j,T}+\Upsilon_{t,T},
\end{eqnarray}
where $X_{1,T},\dots,X_{T,T}$ are observed, $\Upsilon_{1,T},\dots,\Upsilon_{T,T}$ are the innovations with $T$ the sample size. The mean of the observations is given by $\mu_0:=E(X_{t,T})$ and $Z_{t,T}=X_{t,T}-\mu_0$. The dynamics of the series are driven by the autoregressive matrices $A_{0j}$, $j=1,...,p$. Let us denote by $\otimes$ the usual Kronecker product and introduce vec$(.)$ the operator consisting in stacking the columns of a matrix into a vector.
Let the matrix of parameters $\mathbb{A}_0=[A_{01}:\dots:A_{0p}]$. In the sequel we consider the parameters vector $\theta_0=\mbox{vec}\left(\mathbb{A}_0\right)$, so that we can write:

$$Z_{t,T}=(Z_{t-1,T}^p\otimes I_2)\theta_0+\Upsilon_{t,T},$$
with $Z_{t-1,T}^p=(Z_{t-1,T}',\dots,Z_{t-p,T}')$. As pointed out by L\"{u}tkepohl (1982), it may be of interest to add other relevant variables to the system under study.
Here a bivariate system is considered for the sake of conciseness.
The following condition holds throughout the paper although our results can be extended easily to the cointegrated case.\\

\textbf{Assumption A0: Stability of the observed process.}\\
$\det {A}(z)\neq 0$ for
all $|z|\! \leq \! 1$, with ${A}(z) \!=\!
I_d-\sum_{i=1}^{p}\!{A}_{0i} z^i$.\\

We make the following assumption delineating the heteroscedastic structure.\\

\textbf{Assumption A1: Time-varying covariance structure of $(\Upsilon_{t,T})$.}\\ We assume that $\Upsilon_{t,T}=H_{t,T}\epsilon_{t}$ where:\\
(i) The matrices $H_{t,T}$ satisfy $H_{t,T}=G(t/T)$,
and the components $g_{kl}(r)$ of the matrix $G(r)$, $r\in(0,1]$,
are measurable non constant deterministic functions, such that
$\sup_{r\in(0,1]}|g_{kl}(r)|<\infty$. Each $g_{kl}(\cdot)$ satisfies a Lipschitz condition piecewise on a finite number of some sub-intervals that partition $(0,1]$ (the partition may depend on $k,l$). The matrix $G(r)$ is invertible for all $r$.\\
(ii) The process $(\epsilon_t)$ is iid and such that $E(\epsilon_t)=0$, $E(\epsilon_t\epsilon_t')=I_2$,
and $\left(\mbox{E}\|\epsilon_t\|^{4\nu}\right)<\infty$ for some $\nu>1$ and where $\|.\|$ denotes the Euclidean norm.\\

In {\bf A1(i)} we use the rescaling device introduced by Dahlhaus (1997) for defining the $H_{t,T}$'s. This non constant covariance specification is widely used in the literature, and allows for a wide range of unconditional time-varying covariance patterns commonly observed in macroeconomic variables as for instance abrupt shifts or cyclical behaviors. The invertibility assumption on $G(r)$ entails that the $H_{t,T}$'s are invertible, but also that the non constant covariance structure $\Sigma(r)=G(r)G(r)'$ is positive definite for all $r$. The condition {\bf A1(ii)} ensures the identifiability of the covariance structure. Under {\bf A1} an estimator $\hat{\mu}$ such that $\sqrt{T}(\hat{\mu}-\mu_0)=O_p(1)$ is available from Lemma 1 of Xu (2012). Using $Z_{t,T}(\hat{\mu})=X_{t,T}-\hat{\mu}$, it is easy to see from Patilea and Ra\"{\i}ssi (2012) that a $\sqrt{T}$-asymptotically normal estimator of $\theta_0$ is available. In the sequel we denote by $\hat{\theta}$ such an estimator. 
Once the conditional mean is filtered in accordance to the above steps, we can proceed to the analysis of linear relations in the covariance structure. Before this, let us underline that the study of common linear relationships between variance structures is meaningless if the studied processes are in fact homoscedastic. Then it is important to check that the variances are time varying using, for instance, the multivariate CUSUM test proposed by Aue, H\"{o}rmann, Horv\`{a}th and Reimherr (2009).\\

Let us denote by $\Upsilon_{1t,T}$ (resp. $\Upsilon_{2t,T}$) the errors corresponding to the first component of $X_{t,T}$ (resp. second component of $X_{t,T}$), so that we have $\Upsilon_{t,T}:=(\Upsilon_{1t,T},\Upsilon_{2t,T})'$. In this paper we wish to test the null hypothesis ($H_0$) that there exists a linear relationship between the variance structures of $\Upsilon_{1t,T}$ and $\Upsilon_{2t,T}$. More precisely denoting by $\sigma_i^2(r)$ the $i$th diagonal element of $\Sigma(r)$, we test:
\begin{equation}\label{H0}
H_0:\:\sigma_{2}^2(r)=a_0\sigma_{1}^2(r)+b_0,\:\mbox{for all}\:r\in(0,1],
\end{equation}
against the alternative that
$$H_1:\:\sigma_{2}^2(r)\neq a_0\sigma_{1}^2(r)+b_0,\:\mbox{for all}\:r\in[\pi_1,\pi_2],$$
for some fixed $0<\pi_1<\pi_2\leq1$ and $(a_0,b_0)'\in\mathbb{R}^2$.\\

The most popular tests to detect departures from a linear form are the cumulative sum (CUSUM) type tests and the Chow test. Hansen (1992b) has underlined that the CUSUM type tests have several advantages on alternative approaches. In particular CUSUM tests do not need to fix a possibly arbitrary change point on the contrary to the Chow test (see Chow (1960) for more details). This explains why many of the papers dealing with some departure from a null hypothesis through time use CUSUM statistics. Reference can be made to the analysis of non constant variances structures by Cavaliere and Taylor (2007, Section 6) among others. Hence for testing $H_0\:\mbox{vs.}\:H_1$, we consider the CUSUM statistic:

$$S=\sup_{r\in(0,1]}|\delta_r|,$$
where $\delta_r=T^{-\frac{1}{2}}\sum_{t=1}^{[Tr]}\hat{\zeta}_{t,T}$, with $r\in(0,1]$, $\hat{\zeta}_{t,T}=\hat{\Upsilon}_{2t,T}^2-a_0\hat{\Upsilon}_{1t,T}^2-b_0$ and $\widehat{\Upsilon}_{t,T}=(\hat{\Upsilon}_{1t,T},\hat{\Upsilon}_{2t,T})'$ are the residuals of the conditional mean estimation. Now denote by $\sigma_{\xi}^2(t/T)$ the variance of $\zeta_{t,T}:=\Upsilon_{2t,T}^2-a_0\Upsilon_{1t,T}^2-b_0$. The $\sigma_{\zeta}^2(.)$ can be expressed in function of the components of the matrix $G(.)$ and the fourth moments of the components of $\epsilon_t$. The existence of $\sigma_{\zeta}^2(.)$ is ensured by the conditions $\sup_{r\in(0,1]}|g_{kl}(r)|<\infty$ and $\left(\mbox{E}\|\epsilon_t\|^{4\nu}\right)<\infty$ in {\bf A1}.
The following proposition gives the asymptotic behavior of our statistic. Let us denote by $\Rightarrow$ the convergence in distribution.

\begin{proposition}\label{propostu} Under $H_0$ and if the assumptions {\bf A0}, {\bf A1} and model (\ref{model}) hold true, then we have

$$S\Rightarrow\sup_{r\in(0,1]}|L(r)|,$$
where $L(r)=\int_0^r\sigma_{\zeta}(r)dB(r)$ and $B(.)$ is a standard Brownian motion.
\end{proposition}

From the proof of Proposition \ref{propostu}, it is clear that any model for the conditional mean that admits an estimator fulfilling (\ref{eqeqeq}) is suitable to apply the methodology proposed in the paper. On the other hand note that the asymptotic behavior of our test statistic is not standard. For this reason we use the bootstrap methodology to build a feasible test. Note that the block bootstrap (i.e. sampling by blocks) is useless in our case, as it is unable to reproduce the non constant variance structure of the data. For this reason, the commonly used wild bootstrap technique will be considered. For instance Cavaliere and Taylor (2008, 2009) studied the problem of unit root testing in presence of non constant variance in this way. Thus let us define the bootstrap statistics\footnote{We made a slight abuse using the term "bootstrap" as we do not generate bootstrap counterparts of $X_{t,T}$ since we are considering the squares of the residuals.}
$$S^{(b)}=\sup_{r\in(0,1]}\left|\delta_r^{(b)}\right|,\quad\mbox{where}\quad\delta_r^{(b)}=
T^{-\frac{1}{2}}\sum_{t=1}^{[Tr]}\zeta_{t,T}^{(b)},$$
and $\zeta_{t,T}^{(b)}=\hat{\zeta}_{t,T}\eta_t^{(b)}$ with $\eta_t^{(b)}\sim\mathcal{N}(0,1)$, $b=1,\dots,B$, for some large $B$. Let the bootstrap $p$-value

$$p_T^B=\frac{1}{B}\sum_{i=1}^{B}\mathbb{I}\left(S^{(b)}\geq S\right).$$
The null hypothesis is rejected if the bootstrap $p$-value is lower than some predetermined level $\alpha$. The following proposition gives the asymptotic behavior of the bootstrap statistics under $H_0$ and $H_1$. The reader is referred to Gin\'{e} and Zinn (1990) on the convergence concepts in a bootstrap framework.

\begin{proposition}\label{propostu2} Suppose the assumptions {\bf A0}, {\bf A1} and model (\ref{model}) hold true, then we have

\begin{itemize}
\item[(i)] under $H_0$:
$$S^{(b)}\Rightarrow_p\sup_{r\in(0,1]}|L(r)|,$$
where $\Rightarrow_p$ means that weak convergence under the bootstrap probability measure occurs in a set with probability converging to one, and
\item[(ii)] under $H_1$: $p_T^B\rightarrow0,$ in probability as $T\to\infty$.
\end{itemize}
\end{proposition}

The assertion $(i)$ of Proposition \ref{propostu2} ensures that under $H_0$ the bootstrap statistics $S^{(b)}$ have asymptotically the same behavior as that of the original statistic $S$. Under $H_1$, the assertion (ii) ensures that the proposed test is able to detect alternatives. Let us mention that several other distributions for the bootstrap disturbances are available in the literature. The random variables $\eta_t^{(b)}$ may for instance follow the Rademacher or Mammen distributions among others possibilities (see Mammen (1993) for more details on these commonly used distributions). In our empirical size study in Section \ref{secMC} below, we considered standard normal and Rademacher (i.e. $P(\eta_t^{(b)}=-1)=P(\eta_t^{(b)}=1)=0.5$) bootstrap disturbances. However our simulation results show a good control of type I errors for in both cases. This suggests that the specification of the bootstrap disturbances does not change the general conclusions of the paper.

It is interesting to note that the statistic $S$ may be unbalanced in the sense that the maximum value is more likely to be attained in periods where the heteroscedasticity is high, so that the alternatives that occur in periods where the variance is low will be more difficult to detect. In presence of important variance changes one can equally divide the whole sample and apply the above bootstrap procedure on each subsample, rejecting the null hypothesis if it is rejected for at least one subsample. Of course in such a case a Bonferroni correction should be used (see Miller (1981) for the Bonferroni correction technique).

Finally let us say few words on how the $a_0$ and $b_0$ coefficients could be fixed in practice. First the values of the tested coefficients can be suggested from the underlining study, as for example if we test the equality of two variance structures. Also one could consider a grid of possible values. 
On the other hand let us point out that estimating $a_0$ and $b_0$ by OLS for instance lead to a different asymptotic behavior for the test statistic. This asymptotic behavior can be obtained using similar arguments to that of Hansen (1992c). Nevertheless we found that the wild bootstrap procedure is unable to reproduce it. We conjecture that uniform convergence is needed in Proposition \ref{propostu} in such a case.

\section{Numerical illustrations}
\label{numsection}

In this section a simulation study is first conducted. Our methodology is next applied to U.S. regional data.
In the empirical size study, we considered the Rademacher and standard normal distributions for the bootstrap disturbances. Let us recall that the Rademacher distribution is such that $P(\eta_t^{(b)}=-1)=P(\eta_t^{(b)}=1)=0.5$. The test with the standard normal (resp. Rademacher) distribution will be denoted by $Q_{S,norm}$ (resp. $Q_{S,rad}$). The linear relations between variance structures are tested at the nominal level $\alpha=5$\%. In all our experiment we fixed $B=999$. 

\subsection{Monte Carlo experiments}
\label{secMC}

The finite sample properties of the testing procedure introduced above are examined by mean of Monte Carlo experiments. The data generating process (DGP) we use is of the form

\begin{eqnarray*}
X_{t,T}&=&\mu_0+A_{01}Z_{t-1,T}+A_{02}Z_{t-2,T}+\Upsilon_{t,T},\label{modeldgp}
\end{eqnarray*}
with $\mu_0=(0,0)'$ throughout this section, so $Z_{t-1,T}=X_{t-1,T}$. Unless otherwise specified, the mean and autoregressive parameters will be assumed unknown. The autoregressive matrices and the heteroscedastic errors are specified below.

In all our experiments we set
\begin{equation}\label{xequa}
\Upsilon_{t,T}\sim\mathcal{N}(0,\Sigma(t/T)),
\end{equation}
where the components of $\Sigma\left(t/T\right)$ are given by $\Sigma^{11}\left(t/T\right)=1+0.4\exp(t/T)$, $\Sigma^{22}\left(t/T\right)=0.5\Sigma^{11}\left(t/T\right)+\delta f(\frac{1}{0.08}(t/T-0.5))$,
and $\Sigma^{12}\left(t/T\right)=0.8\sqrt{\Sigma^{11}\left(t/T\right)\Sigma^{22}\left(t/T\right)}$ with obvious notations, and $f(.)$ the density of the standard normal distribution. For the empirical size study we take $\delta=0$. In all the cases the process $(\Upsilon_{t,T})$ is independent. The cases $\delta>0$ are considered for the empirical power analysis. Under $H_0$ the covariance setting is inspired by the real data study below. In particular an increasing behavior is found for the variances structures and a high correlation between the errors components is fixed.

We investigate the case where a serial correlation is present in the data leading to a preliminary estimation of the conditional mean. For this purpose VAR(1) ($A_{02}=0_{2\times 2}$) and VAR(2) models are considered. In the case of the VAR(1) model the following AR matrix is used:

$$A_{01}=\left(
           \begin{array}{cc}
             0.9 & -0.1 \\
             0.2 & 0.8 \\
           \end{array}
         \right),
$$
so there is a strong serial correlation, while we consider

$$A_{01}=\left(
           \begin{array}{cc}
             0.5 & 0.2 \\
             -0.1 & -0.4 \\
           \end{array}
         \right),\quad
A_{02}=\left(
           \begin{array}{cc}
             -0.3 & 0.1 \\
             0 & 0.2 \\
           \end{array}
         \right)
$$
in the VAR(2) case.
On the other hand, note that in the methodology described in the previous section, we do not generate bootstrap replicates of the original observations $X_{t,T}$. Hence in order to evaluate the effects of applying the bootstrap directly to the squared residuals, $A_{01}$ and $A_{02}$ are set equal to zero so we have $X_{t,T}=\Upsilon_{t,T}$. Then the results corresponding to the case where the test statistic $S$ is computed from (overfitted) VAR(1) residuals will be compared to those where the $\Upsilon_{t,T}$'s are assumed observed. In the latter case this entails to assume $A_{01}=A_{02}=0_{2\times 2}$ and $\mu_0=(0,0)'$ known, which is not realistic in most of the cases. However recall that we only aim to evaluate the effect of skipping the conditional mean estimation in the bootstrap procedure. In each experiment $N=1000$ independent trajectories are simulated using (\ref{xequa}).\\

First the empirical size of our tests are investigated ($\delta=0$).
Table \ref{tab1} and \ref{tab2} correspond to the case where correct VAR(1) and VAR(2) models are fitted to the simulated data. In Table \ref{tab3} the relative rejections frequencies of the test are given for the case of a mis-specified VAR(1) model fitted to VAR(2) processes. In Table \ref{tab4} the outputs for the independent observed processes are displayed. Since we carried out $N=1000$ independent experiments, and if we suppose that the finite sample size of the tests are $5\%$, the relative rejection frequencies should be between the significant limits 3.65\% and 6.35\% with probability 0.95.\footnote{Using the standard Central Limit Theorem the relative rejection frequencies should be within the confidence interval $\left[5\pm100*1.96*\sqrt{\frac{0.05*0.95}{N}}\right]$ with a probability of approximately 0.95.} Therefore the relative rejection frequencies outside these limits are displayed in bold type.

The relative rejections frequencies in Table \ref{tab1} and \ref{tab2} suggest a good control of the type I error for the proposed test for all the sample sizes. In particular we did not observe clear size distortions that could arise from small samples or from the fact that bootstrap counterparts of $X_{t,T}$ are not generated. However from Table \ref{tab3} it can be seen that the test is oversized when the underlying model is mis-specified, even for large samples. In view of the econometric literature this result is not surprising. For instance Thornton and Batten (1985) or
Stock and Watson (1989) among others have shown that if the underlying VAR structure is not adequately specified, then the linear Granger causality in mean analysis may be quite misleading. Several authors underlined that a correct lag length for VECM forms is crucial for the cointegration analysis (see e.g. Gonzalo and Pitarakis (1998)). In the same way it is important to ensure the goodness-of-fit for the conditional mean before analyzing the non constant variance structure in our case. In the framework of model (\ref{model}), such a task can be carried out using the tools proposed in Patilea and Ra\"{\i}ssi (2013) or Ra\"{\i}ssi (2015).

Next we turn to the outputs for the uncorrelated case. From Table \ref{tab4}, we can see that there is no significant difference between the overfitted case, leading to build the test statistic using residuals, and the case where the $\zeta_{t,T}$'s are assumed observed. Then, it seems that skipping the conditional mean estimation in the bootstrap procedure does not change much the size properties of the test.

Now let us say few words on the choice of the bootstrap disturbances. In general we did not encountered major differences between the outputs of the standard normal and Rademacher distributions (see Tables \ref{tab1}-\ref{tab4}). In particular let us underline that under our hypotheses, and if we suppose that the $\zeta_{t,T}$'s are observed with a symmetric distribution, it can be shown that the bootstrap statistics have exactly the same distribution as $S$ if the bootstrap disturbances follow a Rademacher distribution (see the proof of Theorem 1 in Davidson and Flachaire (2008)). Although the above conditions are met in the left side of Table \ref{tab4}, we do not remark a significant difference between outputs of the Rademacher and standard normal distributions. As a consequence, it seems that the choice of the bootstrap disturbances has a little effect on the empirical size of the test.\\

\begin{figure}[h]
\begin{center}
\includegraphics[scale=0.42]{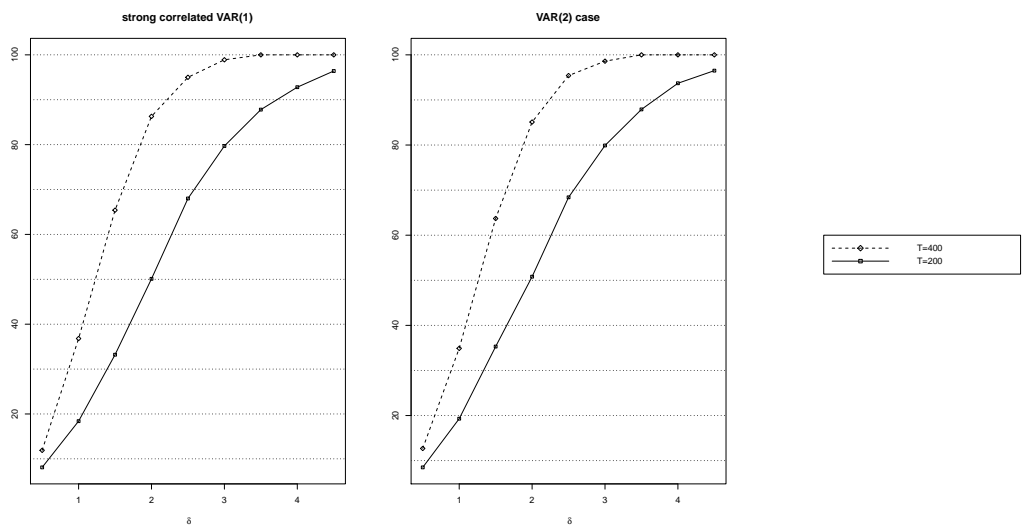}
  \end{center}
\caption{Empirical power (in \%) of the $Q_{S,norm}$ test in the case of the presence of serial autocorrelation. VAR models are adequately adjusted to the data.}
\label{pow1fig}
\end{figure}
\begin{figure}[h]
\begin{center}
\includegraphics[scale=0.60]{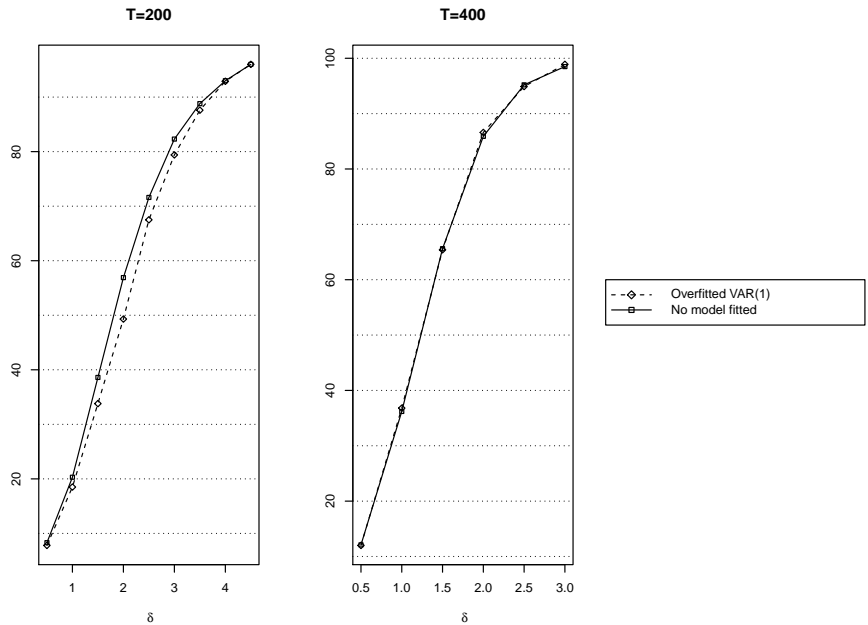}
  \end{center}
\caption{Empirical power (in \%) of the $Q_{S,norm}$ test for uncorrelated simulated processes with mean zero ($A_{01}=0$). The dashed lines correspond to the case where a VAR(1) model with constant is estimated although the simulated process is uncorrelated (the overfitted case). The full lines correspond to the case where no model is fitted to the uncorrelated process.}
\label{pow2fig}
\end{figure}

Now we study the behavior of our test under the alternative. To this aim two samples are considered $T=200,400$, and we take several values for $\delta$. In the sake of conciseness, we only considered the $Q_{S,norm}$ test. Indeed, we also found similar results for the standard normal and the Rademacher disturbances under the alternative hypothesis. In Figure \ref{pow1fig} the relative rejection frequencies in the VAR(1) and the VAR(2) cases are displayed. In each case the conditional mean is correctly specified. The outputs corresponding to uncorrelated $(X_{t,T})$ are given in Figure \ref{pow2fig}. Two situations are examined: the errors are assumed observed and the overfitted VAR(1) case.

The obtained results suggest that the proposed test has a stronger capacity to detect departures from $H_0$ as $T$ is large and $\delta$ is far from zero. When the sample is small ($T=200$), we can also observe some slight loss of power when a conditional mean is estimated in comparison to the case where the $(\zeta_{t,T})$ is assumed observed (see the left panel of Figure \ref{pow2fig}). For larger samples ($T=400$) the right panel of Figure \ref{pow2fig} show that the relative rejection frequencies become similar.\\


In conclusion our experiments show that our test behaves reasonably good for samples sizes commonly observed for heteroscedastic data. On the other hand we did not find major finite sample consequences neglecting the conditional mean estimation step in the bootstrap algorithm or regarding the choice of the bootstrap disturbances.

\subsection{Real data analysis}
\label{realdata}

As an illustrative example we investigate the variance structure of the 4-dimensional system composed by the first differences of the bimonthly U.S. consumer price indexes (CPI) for all urban consumers, all items, from January 1, 1978 to November 1, 2017 for four cities in the U.S.: Chicago, New York, Los Angeles and Boston. The length of the series is $T=239$ after differencing. The data can be downloaded from the website of the research division of the Federal Reserve Bank of Saint Louis. Considering the first differences allow to eliminate the trending behavior in the identification-estimation-validation and forecast Box and Jenkins approach to time series modelling.

\begin{figure}[h]
\begin{center}
\includegraphics[scale=0.60]{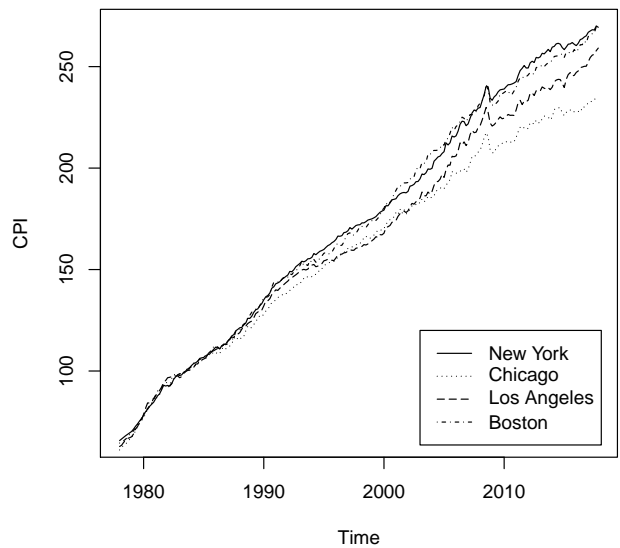}
  \end{center}
\caption{CPI for all urban consumers of New York, Chicago, Los Angeles and Boston from January 1, 1978 to November 1, 2017. Data source: the research division of the Federal Bank of Saint Louis. https://fred.stlouisfed.org/.}
\label{data}
\end{figure}

Regional economics are widely studied in the applied literature. The reader is referred to papers published in specialized reviews as \textit{Journal of Urban Economics} or \textit{Journal of Regional Analysis and Policy} among many others. In view of the plot of the series before computing the first differences, in Figure \ref{data}, a cointegration analysis in a heteroscedastic framework seems relevant (see the tools developed in Cavaliere, Rahbek and Taylor (2010)). In addition as we can clearly observe a global increasing variance, one wonders whether the variabilities of the different CPI's are increasing in the same rate or not. In this way we illustrate how our methodology may serve in finding common patterns in order to make clusters of (regional) time series variables. Note also that clustering is a routine task for time series (see e.g. Shumway (2003) or Hirukawa (2006)).

We adjusted a VAR(4) model to filter out the conditional mean of the studied series. In our VAR system the first component corresponds to the Chicago CPI, the second to the New York CPI, the third to the Los Angeles CPI and the fourth to the Boston CPI.
The model adequacy is checked using the adaptive Box-Pierce portmanteau test proposed in Patilea and Ra\"{\i}ssi (2013). The $p$-value corresponding to 18 autocorrelations in the test statistic is 11.01\%. The existence of second order dynamics in the residuals is tested by considering the Monte Carlo cross validation portmanteau test proposed in Patilea and Ra\"{\i}ssi (2014). The $p$-values corresponding to 1, 3 and 6 autocorrelations, displayed in Table \ref{tsod}, show that a deterministic time-varying variance seems reasonable. Hence our outputs suggest that a VAR(4) model with a deterministic specification for the errors variance structure seems adequate.

Using the residuals obtained from the previous step, we test the following hypotheses, with $p$-values obtained from 999 bootstrap iterations:
\begin{eqnarray*}
E(\Upsilon_{1t,T}^2)&=&E(\Upsilon_{2t,T}^2),\:\mbox{"Chicago=New York"},\: p-\mbox{value}=36.7\%,\\
E(\Upsilon_{1t,T}^2)&=&E(\Upsilon_{3t,T}^2),\:\mbox{"Chicago=Los Angeles"},\: p-\mbox{value}=90.9\%,\\
E(\Upsilon_{1t,T}^2)&=&E(\Upsilon_{4t,T}^2),\:\mbox{"Chicago=Boston"},\: p-\mbox{value}=90.5\%,\\
E(\Upsilon_{2t,T}^2)&=&E(\Upsilon_{3t,T}^2)\:\mbox{"New York=Los Angeles"},\: p-\mbox{value}=9.2\%,\\
E(\Upsilon_{2t,T}^2)&=&E(\Upsilon_{4t,T}^2)\:\mbox{"New York=Boston"},\: p-\mbox{value}=15.7\%,\\
E(\Upsilon_{3t,T}^2)&=&E(\Upsilon_{4t,T}^2)\:\mbox{"Los Angeles=Boston"},\: p-\mbox{value}=85.8\%.
\end{eqnarray*}
Our analysis indicate that there is no evidence to reject an equality relation between the variance shapes of the different CPI's. Note however the variance structure of the New-York CPI seems somewhat away from the other studied cities.

\newpage
\section{Proofs}
\label{secproofs}

\noindent{\bf Proof of Proposition \ref{propostu}.}\quad Define $Z_{t,T}(\mu)=X_{t,T}-\mu$, and $Z_{t-1,T}^p(\mu)=(Z_{t-1,T}(\mu)',$ $\dots,Z_{t-p,T}(\mu)')$ for any $\mu\in\mathbb{R}^2$. Introduce $\Upsilon_{t,T}(\theta,\mu)=Z_{t,T}(\mu)-(Z_{t-1,T}^p(\mu)\otimes I_2)\theta$ for any $\theta\in\mathbb{R}^{4p}$ such that $\Upsilon_{t,T}(\theta,\mu)$ exists.
In particular $\Upsilon_{t,T}(\theta_0,\mu_0)=\Upsilon_{t,T}$ and $\Upsilon_{t,T}(\hat{\theta},\hat{\mu})=\widehat{\Upsilon}_{t,T}$.
Assuming that (\ref{model}) holds true, under {\bf A0}, {\bf A1}, we have from the Mean Value Theorem
{\small\begin{equation*}\label{isapre}
T^{-\frac{1}{2}}\sum_{t=1}^{T}\widehat{\Upsilon}_{it,T}^2=
T^{-\frac{1}{2}}\sum_{t=1}^{T}\Upsilon_{it,T}^2(\theta_0,\hat{\mu})
-\left\{\frac{2}{T}\sum_{t=1}^{T}\Upsilon_{it,T}(\theta,\hat{\mu})R_i Z_{t-1,T}^p(\hat{\mu})\right\}_{\theta=\theta^*}
\sqrt{T}\left(\hat{\theta}-\theta_0\right),\nonumber
\end{equation*}}
where $i=1,2$, $R_1=(1,0)$, $R_2=(0,1)$ and $\theta^*$ is between $\theta_0$ and $\hat{\theta}$. Hence

\begin{equation}\label{eqeqeq}
T^{-\frac{1}{2}}\sum_{t=1}^{T}\widehat{\Upsilon}_{it,T}^2=
T^{-\frac{1}{2}}\sum_{t=1}^{T}\Upsilon_{it,T}^2(\theta_0,\mu_0)+o_p(1),
\end{equation}
since $\sqrt{T}(\hat{\mu}-\mu_0)=O_p(1)$ and $\sqrt{T}(\hat{\theta}-\theta_0)=O_p(1)$. It is clear that $(\zeta_t)$ is independent using {\bf A1(ii)}. On the other hand $E(\zeta_{t,T})=0$ under $H_0$. Then we can apply Theorem 2.1 of Hansen (1992a) writing

$$\delta_r\Rightarrow L(r).$$
The desired result follows from the Continous Mapping Theorem.\\

\noindent{\bf Proof of Proposition \ref{propostu2}.}\quad We skip the proof of the $(i)$ part of the proposition as it follows the same steps of the proof of Lemma A.5 of Cavaliere, Rahbek and Taylor (2010). For the part (ii) of the proposition, note that

{\small\begin{eqnarray*}
T^{-\frac{1}{2}}\delta_r=
T^{-1}\sum_{t=1}^{[Tr]}\zeta_{t,T}+o_p(T^{-\frac{1}{2}})&=&
T^{-1}\sum_{t=1}^{[Tr]}\zeta_{t,T}-E(\zeta_{t,T})+
T^{-1}\sum_{t=1}^{[Tr]}E(\zeta_{t,T})+o_p(T^{-\frac{1}{2}})\\&:=&
F_1(r)+F_2(r),
\end{eqnarray*}}
since (\ref{eqeqeq}) holds under $H_1$. Using Theorem 20.10 in Davidson (1994), $F_1(r)=o_p(1)$ for any $r\in(0,1]$. Introducing $\nu(t/T)=E(\zeta_{t,T})$, we have $\nu(r)\neq0$, for $r\in[\pi_1,\pi_2]$ under $H_1$. Then it is easy to see that

$$\sup_{r\in(0,1]}|F_2(r)|=\sup_{r\in(0,1]}\left|\int_{0}^{r}\nu(s)ds+o_p(1)\right|,
\:\mbox{with}\:\sup_{r\in(0,1]}\left|\int_{0}^{r}\nu(s)ds\right|>0.$$
Hence
\begin{equation}\label{O1}
S=O_p(T^{\frac{1}{2}}).
\end{equation}

Next notice that under $H_1$

$$\delta_r^{(b)}\sim\mathcal{N}\left(0,T^{-1}\sum_{t=1}^{T}\hat{\zeta}_{t,T}^2\right),$$
conditionally on the sequence $\hat{\zeta}_{1,T},\dots,\hat{\zeta}_{T,T}$. On the other hand, from (\ref{eqeqeq}) and using the same arguments to those of the proof of Lemma 1(iii)(a) of Phillips and Xu (2006), it can be shown that

$$T^{-1}\sum_{t=1}^{T}\hat{\zeta}_{t,T}^2\rightarrow C,\:\mbox{in probability},$$
where $C>0$ is a constant. As a consequence, we have

\begin{equation}\label{O2}
S^{(b)}=O_p(1).
\end{equation}
The equations (\ref{O1}) and (\ref{O2}) entail the consistency of our test.

\section*{References}
\begin{description}
\item[]{\sc Aue, A., H\"{o}rmann S., Horv\`{a}th L.
    and Reimherr, M.} (2009) Break detection in the covariance structure of multivariate time series models. \textit{Annals of Statistics} 37, 4046-4087.
\item[]{\sc Cardinali, A., and Nason, G.P.} (2010) Costationarity of locally stationary time series. \textit{Journal of Time Series Econometrics} 2, 1-19.
\item[]{\sc Cavaliere, G., Rahbek, A., and Taylor, A.M.R.} (2010) Testing for co-integration in vector autoregressions with non-stationary volatility. \textit{Journal of Econometrics} 158, 7-24.
\item[]{\sc Cavaliere, G., and Taylor, A.M.R.} (2007) Time-transformed unit-root tests for models with non-stationary volatility. \textit{Journal of Time Series Analysis} 29, 300-330.
\item[]{\sc Cavaliere, G., and Taylor, A.M.R.} (2008) Bootstrap Unit Root Tests for Time Series with Nonstationary Volatility. \textit{Econometric Theory} 24, 43-71.
\item[]{\sc Cavaliere, G., and Taylor, A.M.R.} (2009) Heteroskedastic Time Series with a Unit Root. \textit{Econometric Theory} 25, 1228-1276.
\item[]{\sc Chow, G.C.} (1960) Tests of equality between sets of coefficients in two linear regressions. \textit{Econometrica} 28, 591-605.
\item[]{\sc Chang, T.-H., Huang, C.-M., and Lee, M.-C.} (2009) Threshold effect of the economic growth rate on the renewable energy development from a change in energy price: Evidence from OECD countries. \textit{Energy Policy} 37, 5796-5802.
\item[]{\sc Dahlhaus, R.} (1997) Fitting time series models to nonstationary processes. \textit{Annals of Statistics} 25, 1-37.
\item[]{\sc Davidson, J.} (1994) Stochastic Limit Theory: An Introduction For Econometricians. Oxford University Press.
\item[]{\sc Davidson, R., and Flachaire, E.} (2008) The wild bootstrap, tamed at last. \textit{Journal of Econometrics} 146, 162-169.
\item[]{\sc Engle, R.F., and Granger, C.W.J.} (1987)
Co-integration and error correction: representation, estimation,
and testing. \textit{Econometrica} 55, 251-276.
\item[]{\sc Engle, R.F., Ng, V.K., and Rothschild, M.} (1990) Asset pricing with a factor ARCH covariance structure: empirical estimates for treasury bills. \textit{Journal of Econometrics} 45, 213-238.
\item[]{\sc Gin\'{e}, E. and Zinn, J.} (1990) Bootstrapping general empirical measures. \textit{Annals of Probability} 18, 851-869.
\item[]{\sc Gonzalo, J., and Pitarakis, J.Y.} (1998) Specification via model selection in vector error correction models. \textit{Economics Letters} 60, 321-328.
\item[]{\sc Hansen, B.E.} (1992a) Convergence to stochastic integrals for dependent heterogeneous process. \textit{Econometric Theory} 8, 489-500.
\item[]{\sc Hansen, B.E.} (1992b) Testing for parameter instability in linear models. \textit{Journal of Policy Modeling} 14, 517-533.
\item[]{\sc Hansen, B.E.} (1992c) Tests for parameter instability in regressions with I(1) processes. \textit{Journal of Business and Economic Statistics} 10, 321-335.
\item[]{\sc Hirukawa, J.} (2006) Cluster analysis for non-Gaussian locally stationary processes.
\textit{International Journal of Theoretical and Applied Finance} 9, 113-132.
\item[]{\sc Johansen, S., and Nielsen, M.{\O}.} (2019) Nonstationary cointegration in the fractionally cointegrated VAR model. \textit{Journal of Time Series Analysis} 40, 519-543.
\item[]{\sc Kim, C.S., and Park, J.Y.} (2010) Cointegrating regressions with time heterogeneity.
    \textit{Econometric Reviews} 29, 397-438.
\item[]{\sc Li, W. K., Ling, S., and Wong, H.} (2001) Estimation for partially nonstationary multivariate autoregressive models with conditional heteroscedasticity. \textit{Biometrika} 88, 1135-1152.
\item[]{\sc Lise, W., and Van Montfort, K.} (2007) Energy consumption and GDP in Turkey: Is there a cointegration relationship? \textit{Energy Economics} 29, 1166-1178.
\item[]{\sc L\"{u}tkepohl, H.} (1982) Non-causality due to omitted
variables. \textit{Journal of Econometrics} 19, 367-378.
\item[]{\sc Mammen, E.} (1993) Bootstrap and wild bootstrap for high dimensional linear models. \textit{Annals of Statistics} 21, 255-285.
\item[]{\sc Miller, R.G.} (1981) Simultaneous Statistical Inference. Second edition, Springer New-York.
\item[]{\sc M\"{u}ller, U.K., and Watson, M.W.} (2018) Long-run covariability. \textit{Econometrica} 86, 775-804.
\item[]{\sc Oh, W., and Lee, K.} (2004) Causal relationship between energy consumption and GDP revisited: the case of Korea 1970-1999. \textit{Energy Economics} 26, 51-59.
\item[] {\sc Patilea, V., and Ra\"{i}ssi, H.} (2012) Adaptive estimation of vector autoregressive models with time-varying variance: application to testing linear causality in mean. \textit{Journal of Statistical Planning and Inference} 142, 2891-2912.
\item[] {\sc Patilea, V., and Ra\"{i}ssi, H.} (2013) Corrected portmanteau tests for VAR models with time-varying variance. \textit{Journal of Multivariate Analysis} 116, 190-207.
\item[] {\sc Patilea, V., and Ra\"{i}ssi, H.} (2014) Testing second order dynamics for autoregressive processes in presence of time-varying variance. \textit{Journal of the American Statistical Association} 109, 1099-1111.
\item[]{\sc Phillips, P.C.B., and Xu, K.L.} (2006)
Inference in autoregression under heteroskedasticity.
\textit{Journal of Time Series Analysis} 27, 289-308.
\item[]{\sc  Ra\"{i}ssi, H.} (2015) Autoregressive order identification for VAR models with non-constant variance. \textit{Communications in Statistics: Theory and Methods} 44, 2059-2078.
\item[]{\sc Ramos, S., and Veiga, H.} (2014) \textit{The Interrelationship Between Financial and Energy Markets.} Springer, Berlin.
\item[]{\sc Sensier, M., and van Dijk, D.} (2004) Testing for volatility changes in U.S. macroeconomic time series. \textit{Review of Economics and Statistics} 86, 833-839.
\item[]{\sc Shumway, R. H.} (2003) Time-frequency clustering and discriminant analysis. \emph{Statistics and Probability Letters} 63, 307-314.
\item[]{\sc Stock, J.H., and Watson, M.W.} (1989) Interpreting the evidence on
money-income causality. \emph{Journal of Econometrics} 40, 161-181.
\item[]{\sc Thornton, D.L. and Batten, D.S.} (1985) Lag-length selection and tests of Granger causality between money and income. \emph{Journal of Money, Credit, and Banking} 17, 164-178.
\item[]{\sc Trimbur, T.M., and Bell, W.R.} (2010) Seasonal heteroscedasticity in time series data: modeling, estimation, and testing. In W. Bell, S. Holan, and T. Mc Elroy (Eds.), \emph{Economic Time Series: Modelling and Seasonality}. Chapman and Hall, New York.
\item[]{\sc Wang, S., Zhao, Q., and Li, Y.} Testing for no-cointegration under time-varying variance. \textit{Economics Letters} 182, 45-49.
\item[]{\sc Wu, C.F.G.} (1986) Jacknife, bootstrap and other resampling methods in regression analysis. \textit{Annals of Statistics} 14, 1261-1295.
\item[]{\sc Xu, K.L.} (2012) Robustifying multivariate trend tests to nonstationary volatility. \textit{Journal of Econometrics} 169, 147-154.
\end{description}

\newpage

\section{Tables and Figures}

\begin{table}[hh]\!\!\!\!\!\!\!\!\!\!
\begin{center}
\caption{\small{Empirical size (in \%) in the VAR(1) case. 
}}
\begin{tabular}{|c|c|c|c|c|c|}
  \hline
  $T$ & 100 & 200 & 400 & 800 & 1600 \\
  \hline
  $Q_{S,norm}$ & 5.4& 5.2& 5.4& 3.7&  4.4 \\
  $Q_{S,rad}$ &5.1& 4.2& 5.3& 6.3&  4.5  \\
  \hline
\end{tabular}
\label{tab1}
\end{center}
\end{table}

\begin{table}[hh]\!\!\!\!\!\!\!\!\!\!
\begin{center}
\caption{\small{Empirical size (in \%) in the VAR(2) case.}}
\begin{tabular}{|c|c|c|c|c|c|}
  \hline
  $T$ & 100 & 200 & 400 & 800 & 1600 \\
  \hline
  $Q_{S,norm}$ & 5.4& 5.9& 4.3& 5.4&  5.1 \\
  $Q_{S,rad}$ & 5.9& 5.5& 3.9& 5.3&  {\bf 6.6} \\
  \hline
\end{tabular}
\label{tab2}
\end{center}
\end{table}

\begin{table}[hh]\!\!\!\!\!\!\!\!\!\!
\begin{center}
\caption{\small{Empirical size (in \%) in the case where a VAR(1) model is fitted to VAR(2) processes (mis-specified case).}}
\begin{tabular}{|c|c|c|c|c|c|}
  \hline
  $T$ & 100 & 200 & 400 & 800 & 1600 \\
  \hline
  $Q_{S,norm}$ & {\bf 10.7}&{\bf 15.2}&{\bf 20.7}&{\bf 31.2}& {\bf 57.5} \\
  $Q_{S,rad}$ & {\bf 10.9}&{\bf 14.7}&{\bf 21.9}&{\bf 30.4}& {\bf 55.6} \\
  \hline
\end{tabular}
\label{tab3}
\end{center}
\end{table}

\begin{table}[hh]\!\!\!\!\!\!\!\!\!\!
\begin{center}
\caption{\small{Empirical size (in \%) in the case of uncorrelated observed processes. On the right a VAR(1) is adjusted to the simulated data and then the test is applied to the residuals (the overfitted case). On the left nothing is estimated prior testing the linear relationship between variance structures.}}
\begin{tabular}{|c|c|c|c|c|c||c|c|c|c|c|}
 \hline
  Case: &\multicolumn{5}{|c||}{no model}&\multicolumn{5}{|c|}{VAR(1)}\\
 \hline
  $T$ & 100 & 200 & 400 & 800 & 1600& 100 & 200 & 400 & 800 & 1600 \\
  \hline
  $Q_{S,norm}$ &4.6& 5.9& 4.6& 4.3&  5.5 & 5.8& 4.9& 5.3& 3.6&  5.1 \\
  $Q_{S,rad}$ & 5.1& 4.3& 4.6& 4.5&  4.7 & 5.5& 4.1& 5.6& 6.2&  4.3 \\
  \hline
\end{tabular}
\label{tab4}
\end{center}
\end{table}

\begin{table}[hh]\!\!\!\!\!\!\!\!\!\!
\begin{center}
\caption{\small{The $p$-values (in \%) of the Monte Carlo cross validation tests for second order correlation (see Patilea and Ra\"{\i}ssi (2014)) for the four residual series obtained from the estimation of the VAR(4) model adjusted to the Chicago, New York, Los Angeles and Boston CPI data. We use 1,3 and 6 autocorrelations for the test statistics. }}
\begin{tabular}{|c|c|c|c|}
  \hline
  Autocor. & 1 & 3 & 6 \\
  \hline\hline
  1 & 78.4 & 95.0 & 94.6 \\
  \hline
  2 & 43.8 & 23.0 & 25.0 \\
  \hline
  3 & 92.2 & 99.8 & 97.4 \\
  \hline
  4 & 57.2 & 36.4 & 41.2 \\
  \hline
\end{tabular}
\label{tsod}
\end{center}
\end{table}

\end{document}